\documentclass[aps,prb,twocolumn,letterpaper,10pt]{revtex4-1}
\usepackage[draft]{graphicx}
\usepackage{textcomp}
\usepackage{txfonts}
\usepackage[dvipsnames]{xcolor}
\usepackage{epstopdf}
\usepackage{lineno}

\begin{document}

\title{Local structure of 
relaxor ferroelectric 
Sr$_x$Ba$_{1-x}$Nb$_2$O$_6$ from pair distribution function analysis}

\author{M. Pa\'{s}ciak$^{\rm a}$}
\email{pasciak@fzu.cz}
\author{P. Ondrejkovic$^{\rm a}$}
\author{P. Vanek$^{\rm a}$}
\author{J. Drahokoupil$^{\rm a}$}
\author{G. Steciuk$^{\rm a}$}
\author{L. Palatinus$^{\rm a}$}
\author{T. R. Welberry$^{\rm b}$}
\author{J. Kulda$^{\rm c}$}
\author{H. E. Fischer$^{\rm c}$}
\author{J. Hlinka$^{\rm a}$}
\author{E. Buixaderas$^{\rm a}$}
\affiliation{
$^{\rm a}$
Institute of Physics of the Czech Academy of Sciences\\%
Na Slovance 2, 182 21 Prague 8, Czech Republic\\
$^{\rm b}$Research School of Chemistry, Australian National
  University, Canberra ACT 2601, Australia\\
$^{\rm c}$Institut Laue-Langevin, BP 156, 38042 Grenoble Cedex 9, France\\
}


\begin{abstract}
Neutron pair distribution function analysis and \textit{ab initio} calculations 
have been employed
to study 
short-range correlations in heavily disordered dielectric material Sr$_x$Ba$_{1-x}$Nb$_2$O$_6$ ($x=0.35, 0.5$ and 0.61). 
The combination of methods has been instrumental/fruitful in pinpointing  
main local-structure features, their temperature behaviour and interrelation. 
A rather complex system of tilts is found to be both temperature and Sr-content sensitive with the biggest tilt magnitudes reached at low temperatures and high $x$. Relative Nb-O$_6$ displacements, directly responsible for material's ferroelectric
properties, are shown to be distinct in two octahedra sub-systems with different 
dynamics
and disparate levels of deviation from macroscopic polarization direction. Intrinsic disorder caused by Sr, Ba and vacancy distribution is found to introduce local strain to the structure and directly influence octahedra tilting. These findings 
establish a new atomistic picture of the local structure -- property relationship in Sr$_x$Ba$_{1-x}$Nb$_2$O$_6$.   
\end{abstract}


\maketitle

\section{Introduction}

Modern materials with interesting or superior properties are often based on structures with  intrinsic imperfections. 
Disruptions to long-range order can modify a collective behaviour of interest -- involving e.g. magnetic spins, electric dipoles or phonons -- and thus can provide an engineering handle on the structure-property relationship. Advanced dielectric materials, like perovskite Pb(Zr,Ti)O$_3$ or PbMg$_{1/3}$Nb$_{2/3}$O$_3$-PbTiO$_3$, are a good example here. They involve relatively simple structural frameworks 
filled with chemical disorder to provide structural heterogeneity that couples to polarization.
However, while the importance of 
employing disorder is clear, exact atomic-scale mechanisms that lead to enhanced properties are usually very difficult to pin-point.~\cite{Keen2015,Zhang2018}   

This is exactly the case of Sr$_x$Ba$_{1-x}$Nb$_2$O$_6$ (SBN100$x$ or shortly SBN). It is a material with a range of attractive properties including broad-band second harmonic generation
as well as strong electro-optic,
dielectric, pyroelectric, and piezoelectric effects.~\cite{Lenzo1967,Neurgaonkar1988,Horowitz1993,Glass1969,Gorfman2015}
A considerable advantage of SBN comes from the fact that its high 
dielectric
response to electrical field is not confined to a narrow temperature range and can be fine-tuned by altering the exact composition.~\cite{Glass1969,Lukasiewicz2008}
%
Moreover, Sr-rich compositions of SBN exhibit a broad frequency- and temperature-dependent permittivity typical for relaxor ferroelectrics.~\cite{Buixaderas2005,Lukasiewicz2008}

\begin{figure}[tbp]
\includegraphics[width=85mm,draft=false]{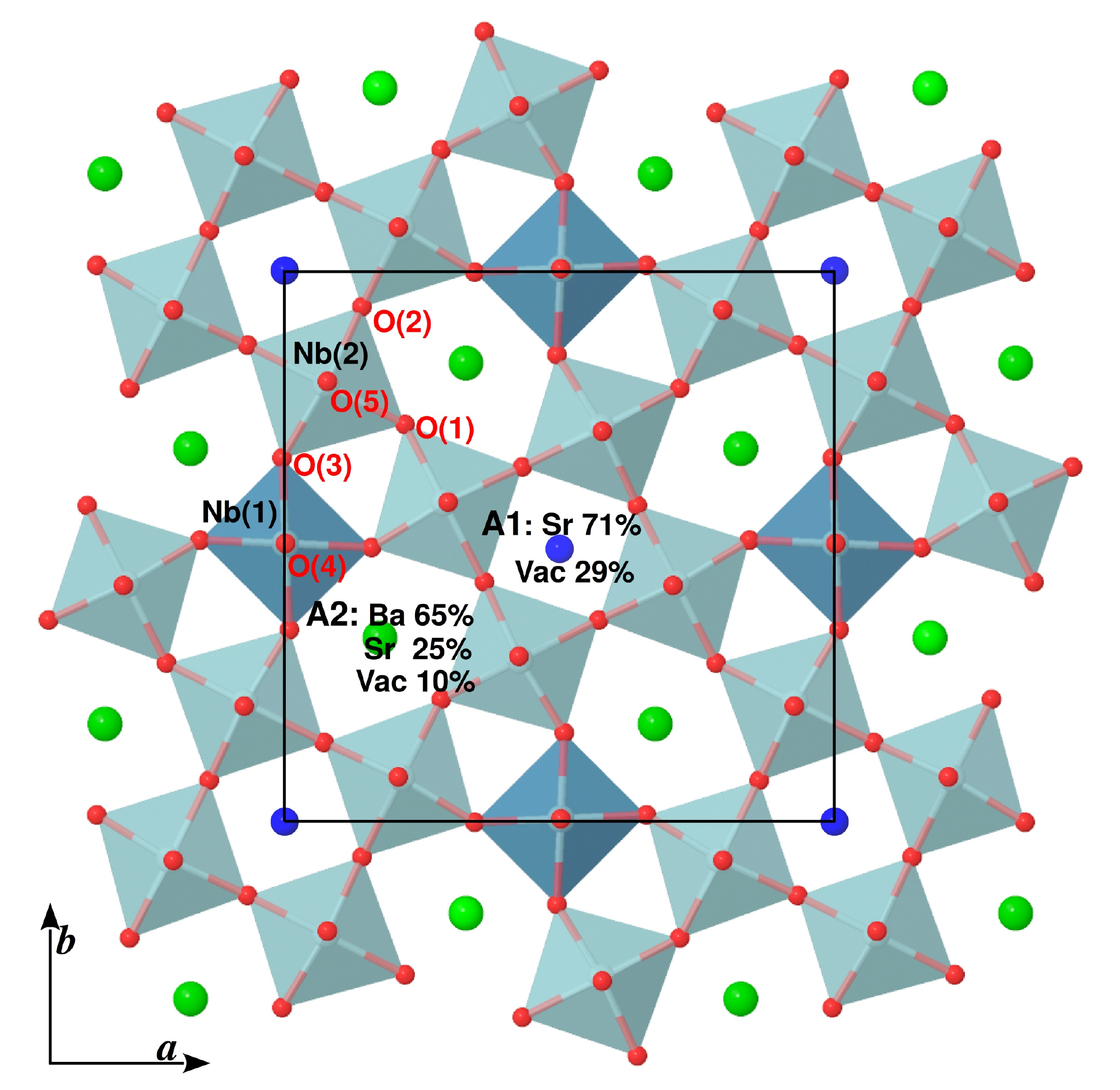}
 \caption{Average structure of SBN48 according to Podlozhenov \textit{et al.}~\cite{Podlozhenov2006} Black square marks 12.484$\times$12.484$\times$3.957~\AA{}$^3$ unit cell. The smallest, triangular channels remain unoccupied; squared A1 channels 
($2a$ Wyckoff sites)
are occupied exclusively by Sr atoms with 12-fold coordination akin to that of A cations in perovskites; and pentagonal A2 channels 
($4c$ Wyckoff sites)
can be occupied by both Sr and Ba atoms with the respective occupation factors. There are two independent Wyckoff positions for the Nb atoms, $2b$ with the site symmetry $mm2$ for Nb(1) and general position $8d$ for Nb(2).
} \label{stru}
\end{figure}

SBN
belongs to the family of compounds with a general formula [(A1)$_2$(A2)$_4$C$_4$][(B1)$_2$(B2)$_8$]O$_{30}$ and 
tetragonal tungsten bronze (TTB) structure in which a network of corner-sharing 
oxygen octahedra creates 
triangular, square, and pentagonal channels running in the direction of the tetragonal $c$ axis (Fig.~\ref{stru}). The TTB structure can be viewed as composed of misaligned, A1B2-based \textit{perovskite-like} units 
infinite in $c$ and comprising four octahedra in the $ab$ plane. In between these units there are single-octahedra,  \textit{linking} columns based on the B1 cation. In the case of SBN, Nb(1) and Nb(2) atoms play the role of B1 and B2 cations, Sr atoms can occupy both the A1 and the A2 positions and large Ba atoms can only fit into the latter. The TTB structure remains unfilled -- the triangular (C) channels are empty and overall occupancies of square (A1) and pentagonal (A2) channels equal $66-72$\,\% and $\sim 90$\,\%, respectively, varying very little with the Sr concentration.~\cite{Podlozhenov2006,Graetsch2017} What does change with $x$, is the share of the sites occupied by Sr and Ba in the A2 channels. The distribution of cations and vacant sites in A1 and A2 channels results in an inherent disorder in SBN.

\begin{figure}[tbp]
\includegraphics[width=62mm,draft=false]{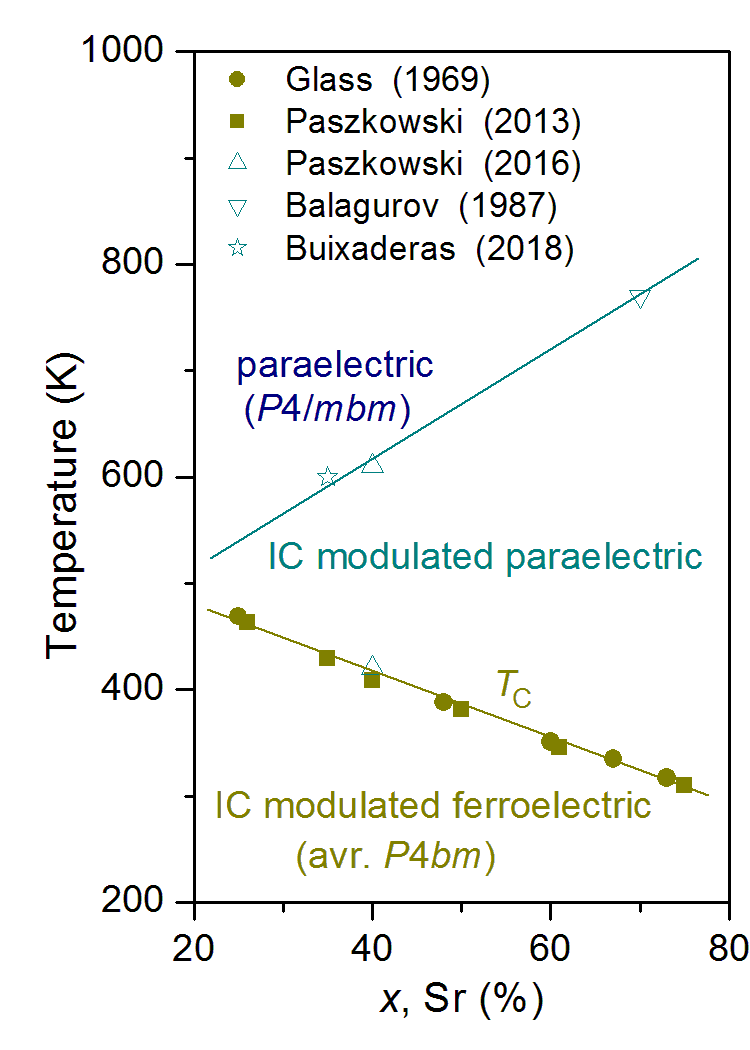}
\caption{Phase diagram of SBN from x-ray diffraction \cite{Paszkowski2013,Paszkowski2016}, electron diffraction \cite{Buixaderas2018}, neutron scattering  \cite{Balagurov1987}, and pyroelectric measurements .\cite{Glass1969}
Characteristic temperatures of distinct temperature regions in relaxor-like compositions are omitted for clarity.
} 
\label{phase diagram}
\end{figure}

The average, room-temperature SBN100$x$ structure for  $x$=0.32-0.82 has been refined 
within the polar $P4bm$ space group with lattice parameters $a$=12.48-12.42~\AA{} and $c$=3.97-3.91~\AA{}.~\cite{Jamieson1968,Podlozhenov2006,Paszkowski2013} However, already an early study by Jamieson \textit{et al.}~\cite{Jamieson1968} revealed substantial anisotropic displacive disorder of Sr/Ba atoms in the A2 channels and a split position for apical oxygen atoms O(4) and O(5). Both features have been later associated~\cite{Bursill1987} with observation of incommensurate diffraction spots with a wave vector ${\bf q} \approx 0.3{\bf a}^\star \pm  0.3{\bf b}^\star + 0.5{\bf c}^\star$.~\cite{Schneck1981} 
The structure was subsequently refined in a (3+2)-dimensional superspace~\cite{Schaniel2002,Woike2003,Schefer2006} with the most pronounced feature 
being a wave-like distribution of NbO$_6$ octahedra tilting 
that superseded the oxygens' split positions. 
While the modulation vector has been shown to change little with composition, the temperature at which satellite spots appear is much higher for bigger Sr content (see Fig.~\ref{phase diagram}, ref.~\onlinecite{Balagurov1987,Paszkowski2016,Buixaderas2018}). Similarly, room-temperature amplitudes of modulation are much bigger for Sr-rich compositions.~\cite{Schefer2008,Graetsch2017}

Ferroelectricity of SBN has been associated with collective Nb displacements from the centers of octahedra in the tetragonal direction.~\cite{Jamieson1968,Olsen2016}
Dielectric studies indicate strong  anomalies with maxima in the temperature range of 310-463~K ($x$=0.75-0.26, Fig.~\ref{phase diagram}) with growing relaxor character of the anomaly for increasing Sr content.~\cite{Buixaderas2005,Lukasiewicz2008,Paszkowski2013}
However, a non-polar high-temperature phase that would see Nb atoms sitting in the octahedra centers has proven somewhat elusive with studies 
showing that $P4/mbm$ space group is adopted at temperatures much higher than $T_{\rm C}$.~\cite{Paszkowski2016,Buixaderas2018}
Moreover, Jamieson et al. proposed an intermediate non-polar and non-centrosymmetric structure to exist above the $T_{\rm C}$ in SBN75.~\cite{Jamieson1968}
Recent structure refinement of a flux grown SBN53 
($T_{\rm C} \approx 368$\,K)
has shown that at 600~K it can be still described in terms of a polar space group $P4bm$.~\cite{Graetsch2014}

These uncertainties can be rationalized taking into account the fact that clusters of polarized NbO$_6$ chains have been shown to exist above the $T_{\rm C}$.~\cite{Shvartsman2008} The chains produce strong diffuse scattering in the form of planes perpendicular to $\mathbf{c}^*$
.~\cite{Bosak2015,Bosak2016,Pasciak2018} 
Several neutron scattering studies of SBN61 showed that this transverse diffuse scattering is present in a broad range of temperatures~\cite{Prokert1978,Gvasaliya2014,Borisov2013,Ondrejkovic2014,Ondrejkovic2016} and carries a signature of growing correlation length and slowing down of clusters across the phase transition.~\cite{Prokert1978,Ondrejkovic2014}

The behaviour of polarization in the presence of disorder in SBN has been considered to conform with the 3D random field Ising model universality class.~\cite{Dec2001} Nevertheless, the atomistic picture that would 
explain the role of vacancies, Sr/Ba substitutional disorder and octahedra tilting is missing. A recent first-principles study~\cite{Olsen2016} of hypothetical SrNb$_2$O$_6$ and BaNb$_2$O$_6$ 
with
the TTB structure went someway towards filling this gap, showing that the distribution of vacancies on A1 and A2 positions substantially changes local energetic landscape and the tendency to develop octahedra tilts or/and polarization.

In this work we contribute to the atomistic description of SBN by studying local structural phenomena in three compositions ($x$ = 0.35, 0.5, and 0.61) over a wide temperature range. As an experimental probe we use the neutron pair distribution function (PDF) method~\cite{Egami2003} that has proven fruitful in the understanding of short-range polar order in perovskite ferroelectric relaxors~\cite{Jeong2005,Laulhe2009,Whitfield2016} and in Pb(Zr,Ti)O$_3$.~\cite{Zhang2014,Zhang2018} The experiment is complemented with \textit{ab initio} calculations that allow us to understand the observed PDF profiles and in addition provide insights into local structural features that do not leave any strong signatures in PDF. 
Moreover, finite-temperature first principles molecular dynamics allows us to calculate time averages and to subtract signal broadening by phonons. In this way we can 
reveal a nanoscale structural change in SBN with temperature. We concentrate on three main aspects of the SBN's local structure: polarization, octahedra tilting and effects of substitutional disorder (including vacancies) and discuss how their interplay can result in SBN's unique properties.


\section{Methods}
\label{sec:exp}
\subsection{Experimental}
The 
SBN
powders with Sr concentration of $x$ = 0.35, 0.5, and 0.61 were prepared using solid-state synthesis. The stoichiometric amounts of high purity reagents SrCO$_3$, BaCO$_3$ and Nb$_2$O$_5$ were mixed and mechanochemically activated by intensive milling in a planetary ball mill (a Fritsch pulverisette 7, ZrO$_2$ crucibles and balls). The resulting powders were annealed twice at 1000$^\circ$C for 12 hours in air, with grinding after each annealing. SBN50 and SBN61 were annealed additionally at 1300$^\circ$C for 3 hours in oxygen. The presence of a single phase having a room-temperature average $P4bm$ structure was confirmed for all samples using powder X-ray diffraction.

The neutron scattering experiment~\cite{PasciakExp} was carried out on the D4c hot neutron two-axis diffractometer~\cite{Fischer2002} at the Institut Laue-Langevin in Grenoble (France). 
The instrument was operated in its standard configuration, using the horizontally flat Cu(220) monochromator (neutron wavelength of $\lambda = 0.5$\,\AA) and the Rh filter to avoid the $\lambda/2$ harmonic contamination. 
The $2\theta$-range of measured diffractograms was $1.5^{\circ}-140^{\circ}$, giving a 
momentum-transfer range $Q \in ( 0.3-23.5)$\,\AA$^{-1}$. 
A standard sealed cylindrical vanadium container of inner diameter of 6.8\,mm was used as a sample holder. The powder-filled part of the container had a height of about 6\,cm.
Diffraction measurements were performed in both a cryostat and a furnace to cover the temperature range of $20 - 600$\,K.
For the sake of subtracting background intensity, diffraction patterns of the empty furnace, the empty cryostat, as well as
for the empty container within both sample environments, were measured.
Neutron powder diffractograms $I(Q)$ were calculated by the \textit{d4creg} program. 
The resulting intensity $I(Q)$ was corrected for background intensity of the empty container and for the Placzek inelastic-scattering effect. 
Subsequently, the pair-distribution function (PDF) was calculated using the \textit{d4fou} program which performs a 
sine Fourier transform of the normalized total scattering function $S(Q)$ as follows
\cite{Egami1991,Proffen2009}
\begin{equation}
G(r) = K \frac{2}{\pi} \int_{0}^{Q_{\rm max}} 
[S(Q)-1] Q \sin(Q r) dQ,
\label{eq:pdf}
\end{equation}
where $r$ is is the inter-atomic distance and $K$ is a factor which allows for the absolute
scale normalisation that was confirmed by comparing the slope of the low-r
part of $G(r)$ with the known number density of atoms for each sample. 
Here, we assumed for simplicity that the scattering intensity $I(Q)$ is proportional to $S(Q)$ without making additional corrections for, e.g. absorption, multiple scattering, other inelastic scattering beyond the Placzek correction and the sample shape.

\subsection{\textit{Ab initio} calculations}
\textit{Ab initio} density functional theory-based calculations were performed with the SIESTA code~\cite{Soler2002} using the pseudopotential plane-wave method. The parametrization of Wu-Cohen~\cite{Wu2006} of the
exchange-correlation functional within the general gradient approximation was used,
together with norm-conserving Trouiller-Martins pseudopotentials. The following electronic states
were included: 2s, 2p and 3d for O, valence 5s, 4d, 5p and semicore 4s and 4p for Sr, 5s, 5p, 4d and
semicore 4s and 4p for Nb as well as 6s, 5d, 6p and semicore 5s and 5p for Ba. A k-grid cutoff of 10~\AA{} and a mesh cutoff of 150 Ry were used.

To allow for the distance range of interest to be properly sampled, we chose a relatively regular and large 2$\times$2$\times$8 supercell (with $a=b=12.48$~\AA{} and $c=3.95$~\AA{} of the $P4bm$ unit cell). The Ba and Sr 
occupation factors are chosen to be as close as possible to the experimental values
and are the following: A1 channels -- $75$~\% Sr, $25$~\% vacancy, A2 channels -- $62.5$~\% Ba, $25$~\% Sr, $12.5$~\% vacancy. The distribution of atoms was selected in a quasi-random manner with the only condition being that for each of the 16 1$\times$1$\times$2-building blocks the occupation ratio is preserved. An initial structure had atoms placed according to a SBN48 average structure from Ref.~\onlinecite{Podlozhenov2006} with the exception of Nb atoms that are not displaced in the $c$ direction from the octahedra centers. 

As a first step we used a conjugate gradient (CG) method to optimize atomic positions such that forces acting on atoms were smaller than 0.1~eV/\AA$^3$. To account for the changes of the structure with temperature we used \textit{ab initio} molecular dynamics (MD) as implemented in SIESTA.
We first pumped into the system in a step-wise manner a kinetic energy equivalent to 1000 K. 
Due to a relatively large size of the supercell, long simulation times were not feasible. Nonetheless, we allowed the system to equilibrate at 800~K for 1~ps, then lowered the MD temperature in a step-wise manner down to 50~K. In the last step we performed CG optimization again arriving at a slightly lower energy (12~meV/formula unit) than in the case of the initial optimization. The optimized volume differs very little (0.25~\%) from the experimental one at 300~K as reported by Podlozhenov et al.~\cite{Podlozhenov2006} If we consider average volume at 300~K in the MD calculation, it is 1.1~\% larger than the experimental one.

\section{Experimental results}

\begin{figure}[tbp]
\includegraphics[width=90mm,draft=false]{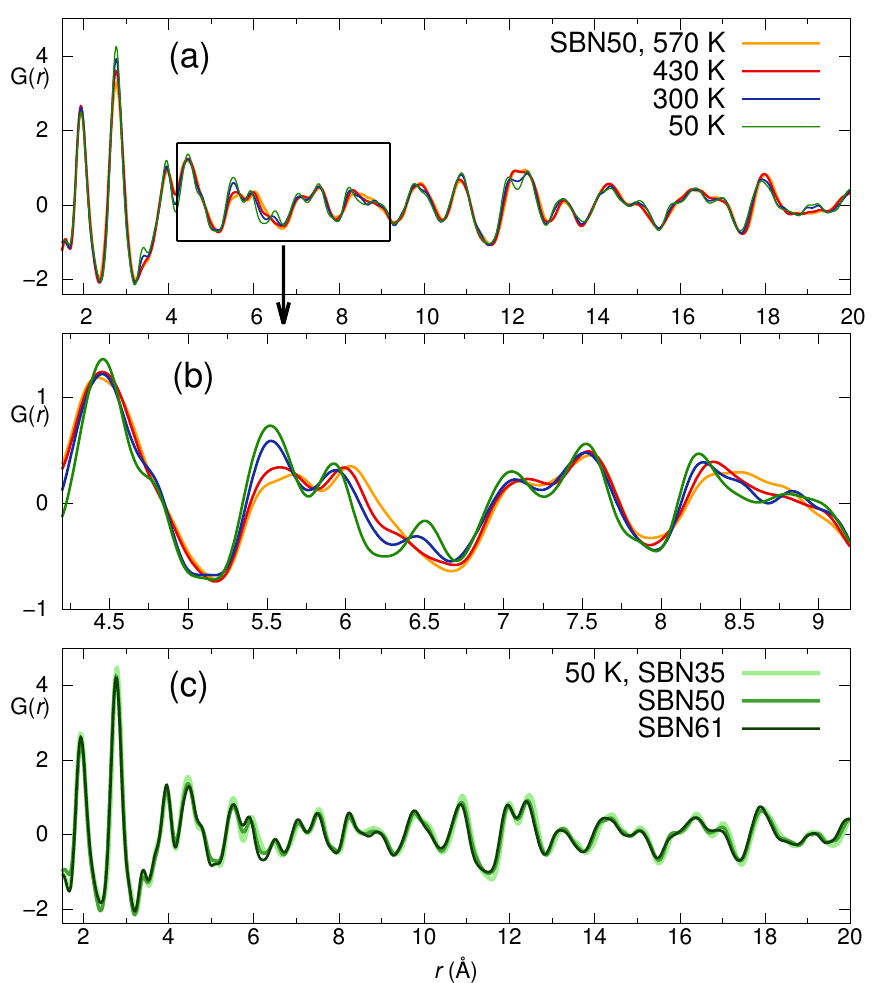}
 \caption{Neutron pair distribution function G($r$) results of SBN100$x$. (a) Temperature dependence of G($r$) for SBN50, (b) detail of $G(r)$ showing a range of distances with the most pronounced changes, (d) full G($r$) of SBN35, SBN50 and SBN61 at 50~K.  
 } 
\label{fig1}
\end{figure}

A
selection of the collected data is presented in Fig.~\ref{fig1}. A remarkable feature characteristic to all $G(r)$ functions is the 
dominating height of the first two peaks (coming from Nb-O and O-O pairs) 
which 
indicates that the main building units -- NbO$_6$ octahedra -- are rather rigid, but their distribution over larger distances is less well defined.
The change of PDF for SBN50 with the temperature in the range of $50 - 570$~K 
($T_{\rm C} \sim 381$~K~\cite{Paszkowski2013})
is presented in Fig.~\ref{fig1}(a). While the positions of most peaks remain unchanged, 
some decrease of heights and broadening can be observed at higher temperatures which is commonly ascribed to increasing thermal motion amplitudes. 
This is especially evident for the O-O peak at $\sim$2.77~\AA{}.
Interestingly, the Nb-O peak with the maximum at $\sim$1.95~\AA{} shows different behavior with smaller temperature dependence.

\begin{figure}[tbp]
\includegraphics[width=80mm,draft=false]{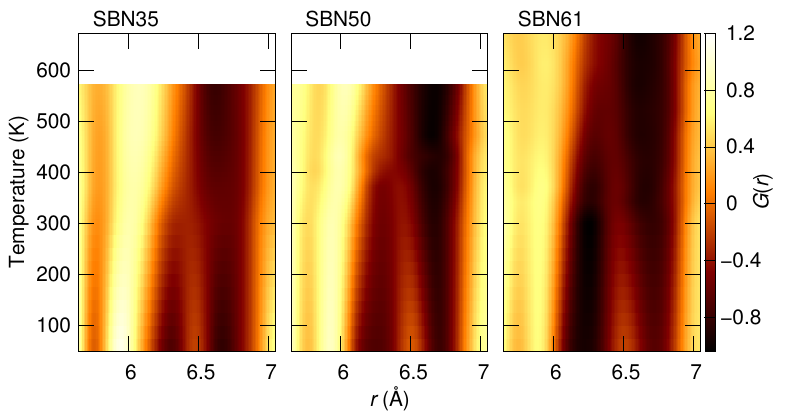}
 \caption{Temperature evolution of G($r$) for SBN35, SBN50, and SBN61 powders. The interpolated maps showing emergence of a peak at $\sim 6.5$\,\AA. 
 } \label{fig2}
\end{figure} 

\begin{figure}[tbp]
\includegraphics[width=90mm,draft=false]{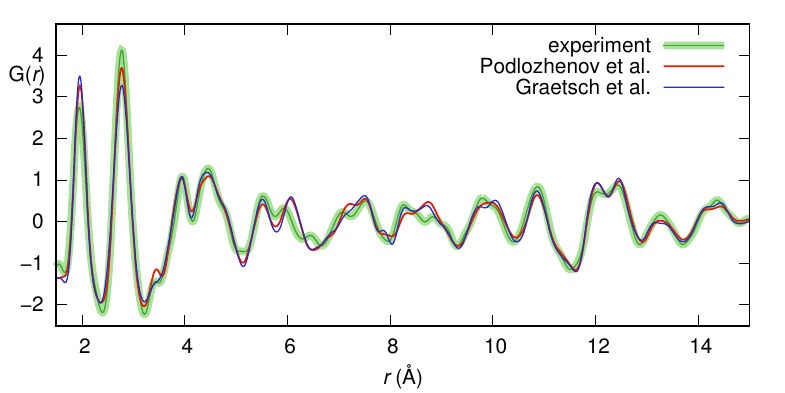}
 \caption{Comparison of room-temperature PDF's: experimental and calculated from an average $Pb4m$ structure of SBN48 (Podlozhenov  \textit{et al.}~\cite{Podlozhenov2006}) as well as from modulated structure of SBN52 (Graetsch \textit{et al.}~\cite{Graetsch2017b}). 
} 
\label{comp_av}
\end{figure} 
The most pronounced temperature dependence is observed at distances $5.25-6.75$~\AA{} and $8.25-9$~\AA{} [enlarged in Fig.~\ref{fig1}(b)] where the height of the peaks is heavily reshuffled and an additional peak emerges at $\sim$6.5~\AA{} on cooling. For SBN50 this peak 
appears at temperatures between $400-500$~K (see Fig.~\ref{fig2}), while its separation begins at higher temperatures for SBN61 and lower for SBN35 clearly implying Sr-content dependence.
In  Fig.~\ref{fig1}(c) the $G(r)$'s for 
the three studied compositions at 50~K are compared showing that the difference between them is rather small, nevertheless it is the biggest for the distances that are the most temperature-dependent. The peak at $\sim$6.5~\AA{} is clearly the largest for SBN61. At the same time most of the PDF peaks get wider with increasing Sr content confirming the more disordered character of Sr-rich SBN.

As a first step towards understanding the experimental PDF we compare it with the $G(r)$ profiles obtained for known structure solutions. We take into account two structures refined from data collected at room temperature: SBN48 average solution within $P4bm$ spacegroup~\cite{Podlozhenov2006} and SBN52 modulated structure.~\cite{Graetsch2017b} In the case of the latter solution a 3$\times$3$\times$1 approximant created with the program Jana2006~\cite{Petricek2014} has been used. We used the program \textsc{pdffit}~\cite{Proffen1999} to refine non-structural parameters (scale, dampening and peak-broadening factors) and get closest possible fits to the experimental data. These are presented in Fig.~\ref{comp_av}. As a first observation we note that the two refined profiles are very similar to each other. They both agree with some parts of the experimental PDF while substantially deviating at certain distances, especially between $5-10$~\AA{}, with the worst fit for the values of $r$ for which the highest temperature sensitivity is observed. 
Additionally a discrepancy in peak widths and heights is observed for the Nb-O peak at 1.95~\AA{} and the peak at 2.77~\AA{} for which the main contribution comes from O-O distances within an octahedron. The first peak is sharper than in the experiment, while it is the opposite in the case of the second, indicating that both models underestimate the spread of Nb-O$_6$ displacements and overestimate  distortion of the octahedra. 

\section{Results of first-principle calculations} 

The local structure of SBN is inevitably affected by the Sr/Ba occupational disorder as well as by the presence of vacancies. This fact together with rather broad features of the PDFs 
suggest that a direct fitting of a model to the collected data is a complicated endeavor. 
We turn, therefore, to first principles calculations which allow for a model-free understanding of our data and facilitate a comprehensive analysis of the nanoscale structure.

\begin{figure}[!tbp]
\includegraphics[width=80mm,draft=false]{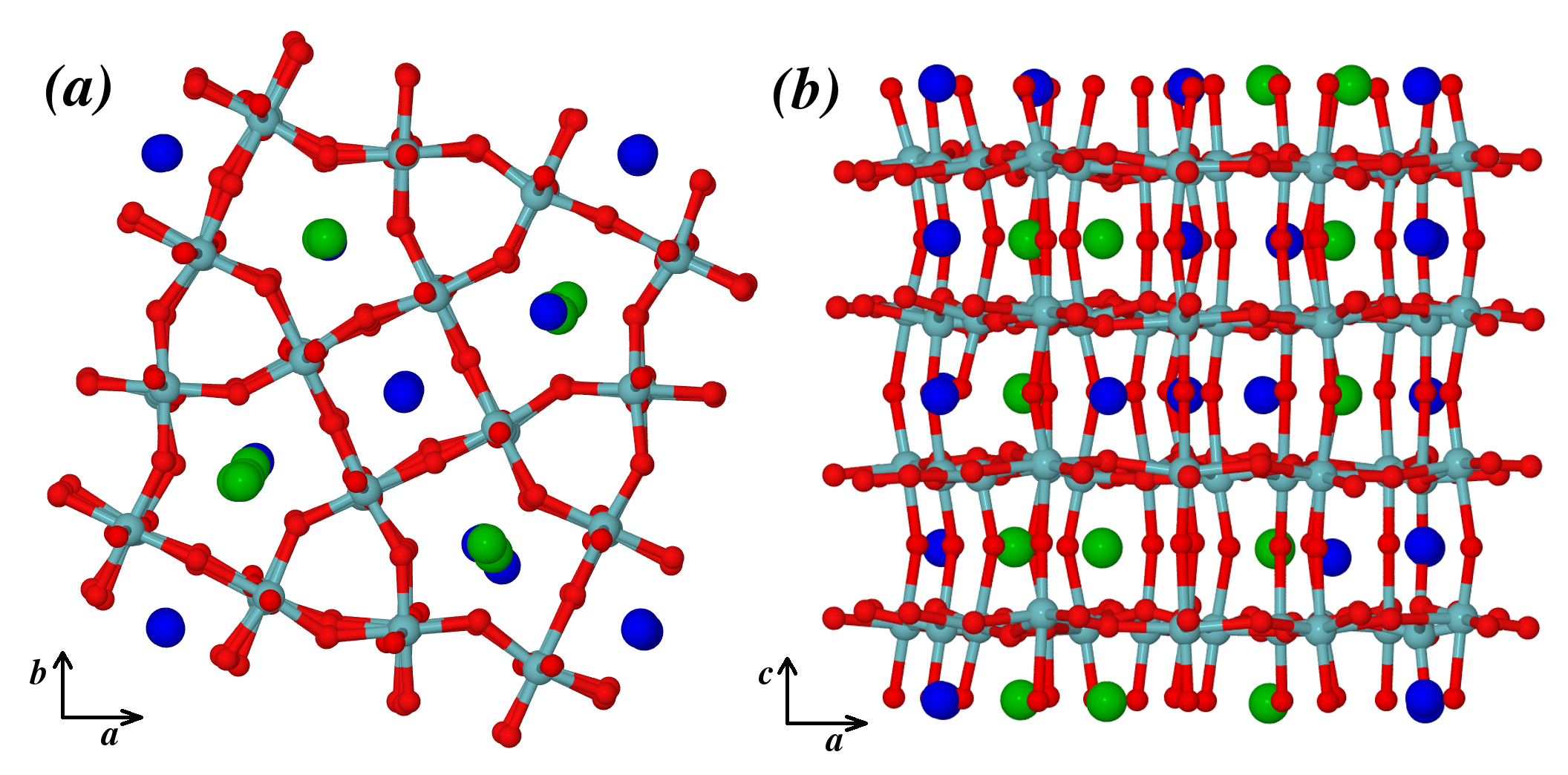}
 \caption{A part of the optimized 2x2x8 supercell viewed along the $c$ direction (a) and $b$ direction (b). Oxygen, niobium, strontium and barium atoms are represented by red, light blue, dark blue and green color, respectively. While displacive disorder in pentagonal channels is evident from (a) with slightly larger displacements for Sr atoms, picture in (b) clearly shows octahedra tilts.
}\label{cg_str}
\end{figure}

The fragment of the optimized structure presented in Fig.~\ref{cg_str} 
reveals important aspects of the static local structure:
Ba/Sr cations are displacively disordered in the pentagonal channels with a large in-plane component of the displacement along clearly preferred directions [cf. Fig.~\ref{cg_str}(a)].
We note that of the two species, Sr atoms tend to shift more from the symmetric $4c$ positions. The structure viewed along the $b$ direction in Fig.~\ref{cg_str}(b) reveals substantial tilting of the oxygen octahedra and, on a closer inspection, a system of alternating shorter and longer Nb-O bonds along the $c$ direction (polarized Nb-O chains).

\begin{figure}[tb]
\includegraphics[width=90mm,draft=false]{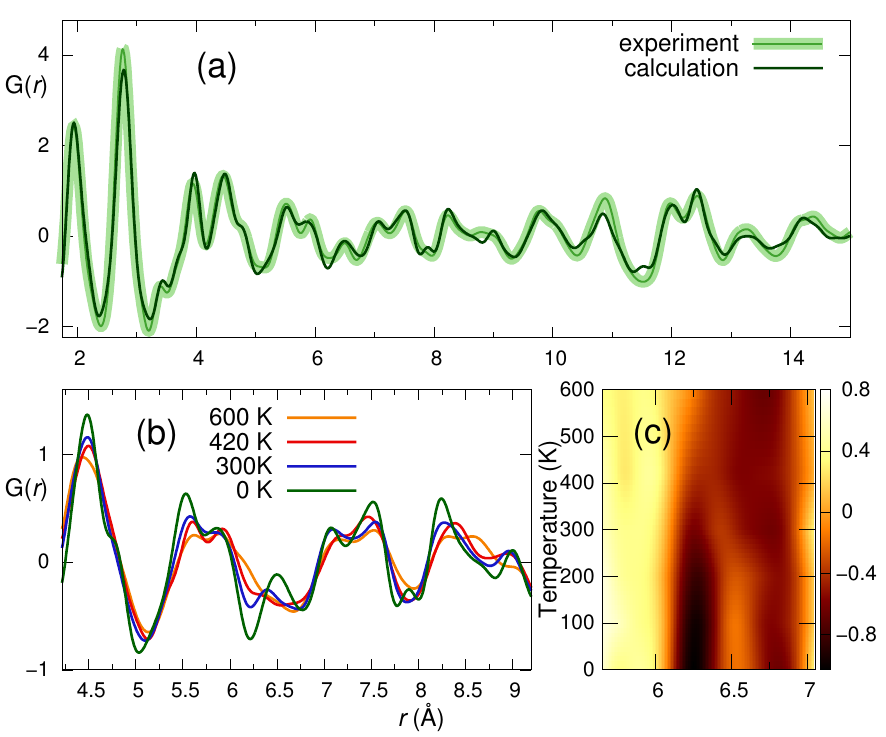}
 \caption{(a) Comparison of experimental (50~K) and calculated (0~K) PDF's for SBN50. The latter data has been corrected for the deviation from the experimental volume (1~\%). No fitting procedure has been involved. (b) Temperature-dependent PDF from \textit{ab initio} molecular dynamics calculations, for comparison with the data presented in Fig.~\ref{fig1}(b). (c) An interpolated map for comparison with experiment in Fig.~\ref{fig2}.} \label{calc_exp_pdf}
\end{figure}


Neutron PDF's from the structures obtained from first principles were calculated with the program DISCUS.~\cite{Proffen1997} In Fig.~\ref{calc_exp_pdf}(a) we present a comparison of the experimental and calculated $G(r)$ profiles which display a remarkable agreement, especially in the previously discussed problematic range of distances ($5-9$~\AA).
To further substantiate the relevance of our calculations, we concentrate on this range and plot $G(r)$ 
obtained from \textit{ab initio} MD calculations at different temperatures
[see Fig.~\ref{calc_exp_pdf}(b)]. We confirm that the same temperature-dependence is observed as in Fig.~\ref{fig1}(b) with an emergence of the peak at $\sim$6.5~\AA{}. 
For a direct comparison with experimental data in Fig.~\ref{fig2}, the temperature-distance map is plotted in Fig.~\ref{calc_exp_pdf}(c).
In view of the good agreement with the experimental results we move to the analysis of the local phenomena in the calculated structures. We will use the PDF 'language' as a convenient way to describe these phenomena even if the direct comparison with the experiment is not possible, e.g. for the partial G(r).    

Let us have a look at the temperature dependence of $G(r)$ from a different angle. Our time-dependent \textit{ab initio} MD calculations can be used to derive time-averaged structures at a given temperature. In this way we can subtract 
 the phononic contribution to the PDF (with a low-frequency limit defined by the simulation time) and concentrate on 
less thermally disordered configurations. In Fig.~\ref{calc_pdf}(a) and (b) we present PDF's of 800~K structures calculated in two different ways and compare them to the fully optimized structure at 0~K. $G(r)$ in Fig.~\ref{calc_pdf}(a) is calculated from a set of instantaneous structures [we average several G(r)'s] representing what would be observed in the experiment. 
Again, the obtained temperature behavior agrees very well with the experimental observations: the first Nb-O peak is virtually temperature-independent, the intensity of the second, O-O peak, changes substantially, and the most pronounced 
changes in the $G(r)$ 
are observed around 6.5~\AA{} and 8.75~\AA{}. 

The 'phonon-filtered' $G(r)$ of the time-average structure
over 1~ps
at 800~K is shown in Fig.~\ref{calc_pdf}(b). 
It allows one to immediately discriminate 
these differences between the 800~K and 0~K $G(r)$ profiles in Fig.~\ref{calc_pdf}(a)  
that are exclusively due to phonon smearing of the signal at high temperatures. For example a small peak at ~3.5~\AA{} or a shoulder at 4.8~\AA{} are both recovered in the 800~K profile upon filtering. Similarly, the O-O peak broadening
in Fig.~\ref{calc_pdf}(a)
can be to a large extent attributed to octahedra distortions due to thermal vibrations.
On the contrary, Nb-O distribution changes from a relatively symmetric peak at high temperatures
in Fig.~\ref{calc_pdf}(b), indicating that Nb atoms sit on average in the center of the octahedra, to an asymmetric peak representative of either dynamic or static off-centering. Again, the most spectacular difference is observed in the range of $5.25-6.75$~\AA{} with an indication that a peak at 6.2~\AA{}, characteristic for the time-averaged structure at high temperatures, heavily splits at low temperatures. The very same 6.2~\AA{} peak is present in the PDF's calculated from the average and from the modulated structures in Fig.~\ref{comp_av}, meaning that the models better represent the average high-temperature structure in this distance region.

\subsection{Octahedral tilts}

\begin{figure}[tbp]
\includegraphics[width=80mm,draft=false]{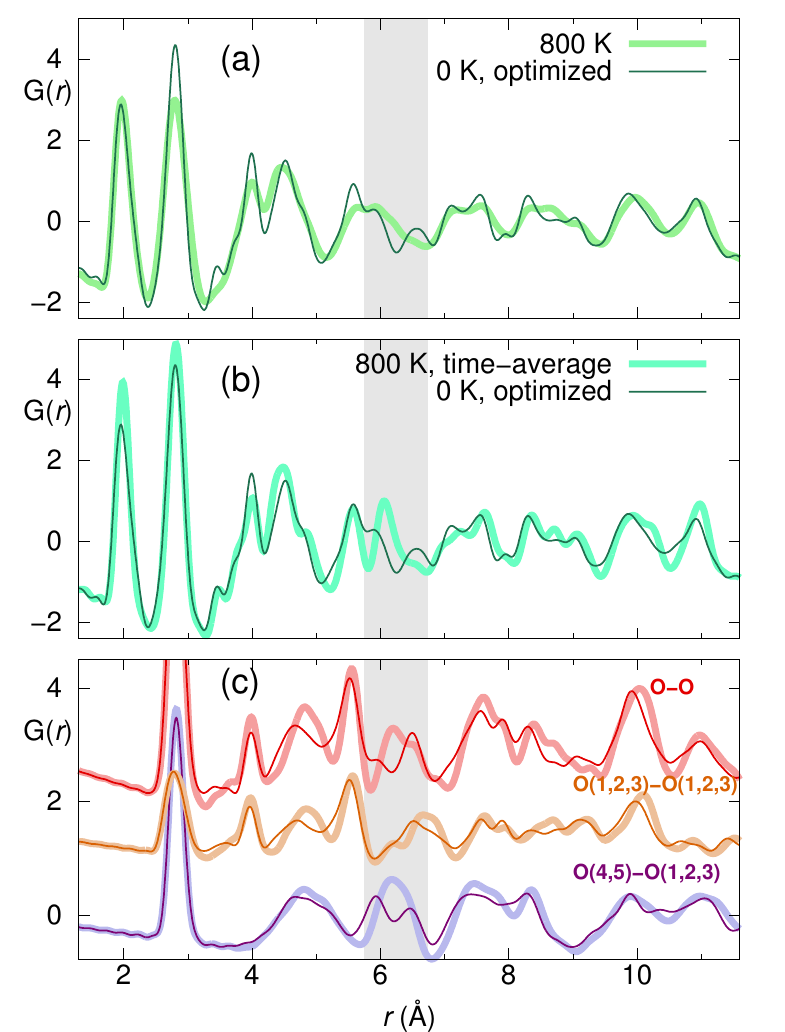}
 \caption{G($r$) 
derived from \textit{ab initio} structures 
of SBN50. 
 (a) Comparison of PDF's for 0~K, CG-optimized (thin line) and 800~K structures (thick line), 
 (b) similar comparison with the 800~K PDF being calculated from the time-averaged structure, 
 (c) partial contributions to PDF's in (b) from all oxygen atoms (top, shifted +2), 
from equatorial oxygen atoms O(1,2,3) 
(middle, shifted +1) and from O(1,2,3) and apical O(4,5) oxygen atoms (bottom, magnified 2x for better comparison). 
See the notation of oxygen atoms in Fig.\,\ref{stru}.
  } \label{calc_pdf}
\end{figure}

\begin{figure}[tbp]
\includegraphics[width=70mm,draft=false]{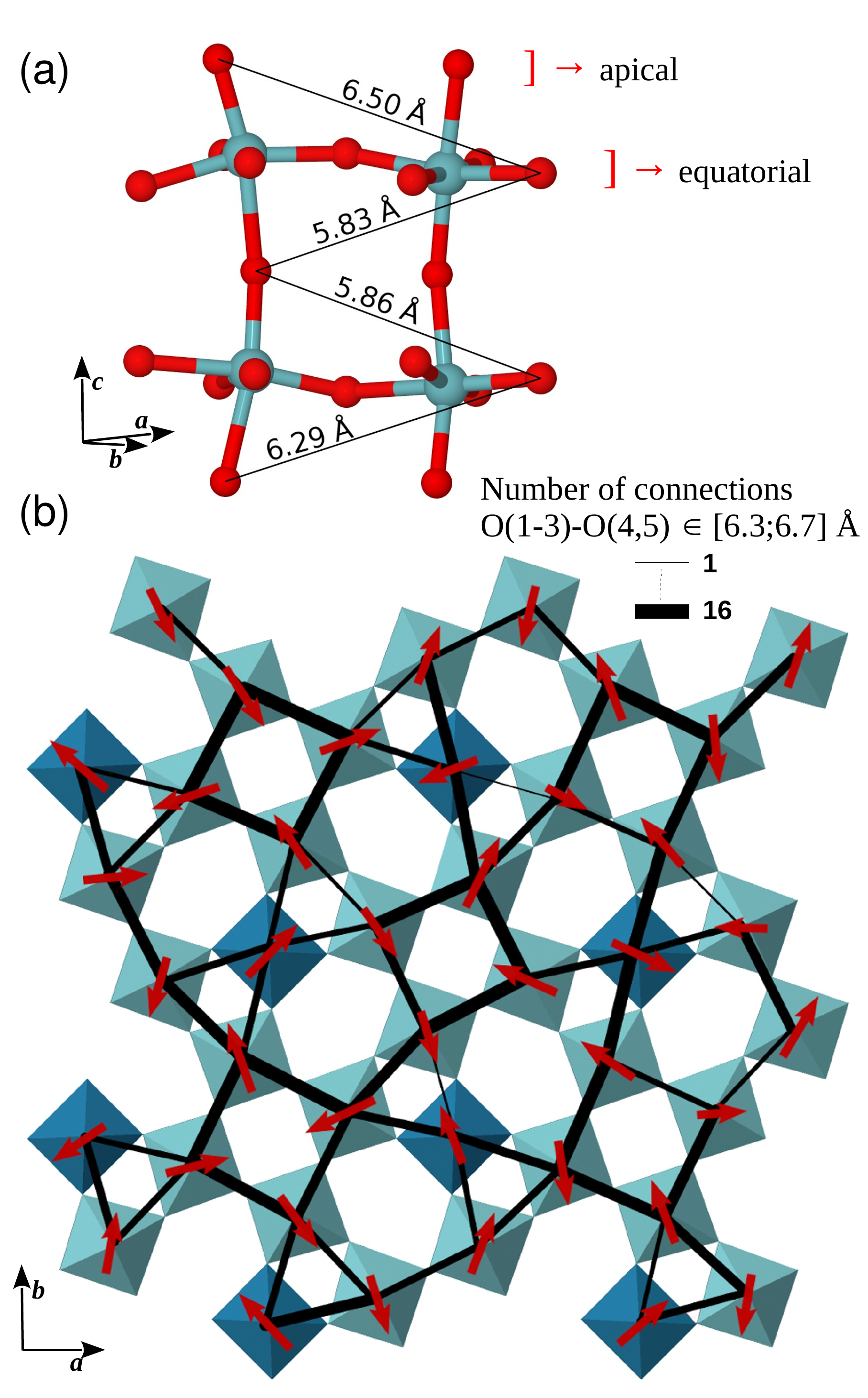}
 \caption{(a) Tilted octahedra with the distances that contribute to the split peak at 6.2~\AA{} in the PDF pattern in Fig.\ref{calc_pdf}. One could draw similar elongated and shortened connections between the neighbouring octahedra in the $c$ direction. (b) Information on octahedral tilts and their contribution to peak splitting integrated over 8 layers. Red arrows correspond to average tilt vectors in one column with every second octahedron taken with an opposite sign to account for the alternating sense of tilt vectors in the $c$ direction. Black lines and their thickness mark the number of connections between apical and equatorial oxygen atoms that fall in the range $6.3-6.7$~\AA{} with the thickest lines having limiting 16 connections.
} \label{tilt}
\end{figure}

From previous studies showing that the main deviation from the average structure 
at lower temperatures 
comprises modulated octahedral tilting, 
we expect that the split at 6.2~\AA{} is a
manifestation of tilts. A confirmation of this can be found in Fig.~\ref{calc_pdf}(c), where partial PDF's are shown for oxygen-oxygen atomic pairs, again calculated in a 'phonon-filtered' manner. If all oxygen atoms are taken into account, we observe a clear emergence of a separate peak at ~6.5~\AA{}. It is convenient to consider two sets of oxygen atoms: 
'apical' ones sharing the $ab$ planes with Sr/Ba atoms [marked as O(4) and O(5) in Fig.~\ref{stru}] 
and 
'equatorial' ones sitting on the same $ab$ planes as Nb atoms 
[marked as O(1), O(2) and O(3) in Fig.~\ref{stru}].
Further partition of $G(r)$ into distributions of 
equatorial-only and apical-equatorial O-O pairs 
reveals 
that the latter case features clear splitting of a broad peak in the distance range of interest.
The way this split is related to tilting can be easily understood from Fig~\ref{tilt}(a), where two octahedra of the 0~K-optimized structure are presented with marked distances that without the tilt are equal.  The fact that this distance partition
is clearly seen in the PDF is due to a combination of at least two factors. First, there is no other strong signal around $\sim$6.5~\AA{}, so at least one of the new peaks can be tracked. Second, the geometry of the octahedral network is such that the relevant oxygen atoms shift in a concerted manner either away from or towards each other creating large difference between the shorter and the longer distances. 

To better understand the nature of the tilts in our \textit{ab initio} optimized structure we plot in Fig.~\ref{tilt}(b)
column-averaged rotation axes (red arrows) represented as vectors with their magnitude proportional to the rotation angle. 
A similar picture is given in the Appendix (Fig.~\ref{tilt_supp}) with the axes being not averaged to show that the dispersion of directions within columns is small.  
We overlay the tilt axes distribution with a network of connections between O(1,2,3) and O(4,5) oxygen atoms in neighbouring columns that fall in the range of 6.3-6.7~\AA{}. 
Again, the information is integrated along the $c$-direction and the thickness of black lines is  proportional to the number of connections between a given pair of columns. In this way we are able to visualize the short-range nature of tilt correlations. Obviously, a 'stronger' connection shows up for octahedra that are tilted towards each other (their rotation axes have a large component transverse to their connecting vector).
One can appreciate that 
Nb(2)O$_6$
octahedra forming square channels (perovskite-like units) have a strong tendency for a correlation of tilts among themselves. 
On the other hand the 'linking' Nb(1)O$_6$ octahedra 
have 'weaker' 
connections, 'transmitting' correlations in selected directions only and creating in this way local tilt paths.
Another interesting observation is the tendency for ordering of tilt axes' directions for 'linking' 
octahedra [darker colour in Fig.~\ref{tilt}(b)]. They are aligned in $[110]$ and alternate in $[\bar{1}10]$, which makes the two diagonal directions inequivalent, clearly breaking the average tetragonal symmetry.   

The average tilt magnitude for the 0~K structure is 8.4(5)\textdegree{}. In fact, the tilts do not disappear in our calculations even at the highest temperatures with their average magnitude being $\sim$5\textdegree{} at 800~K. However, this smaller magnitude accompanied with a moderate deterioration of tilt correlations suffice for the two separate peaks in the PDF to merge into a single broad one.


\subsection{Local polarization}

\begin{figure}[tbp]
\includegraphics[width=90mm,draft=false]{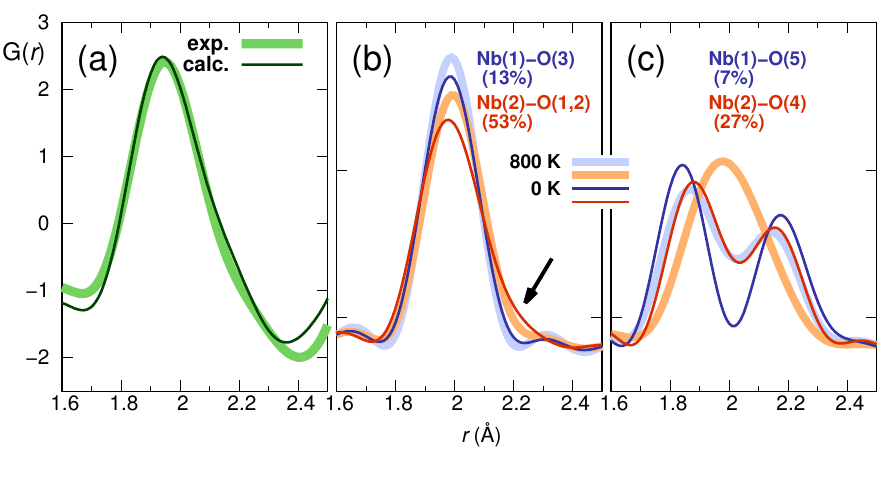}

\includegraphics[width=90mm,draft=false]{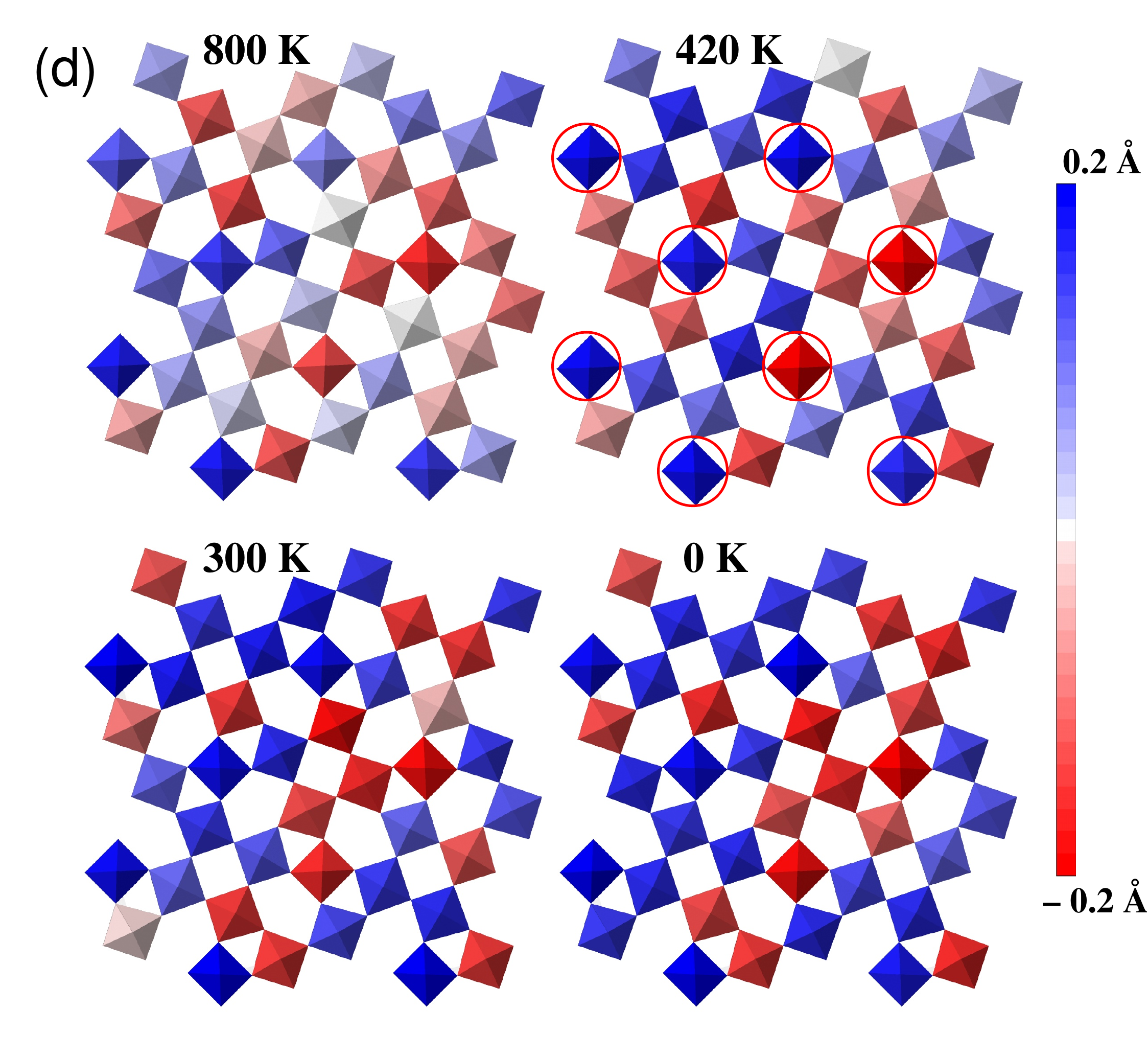} 
 \caption{Detailed analysis of the Nb-O distances within octahedra. 
(a) Comparison of experimental and calculated Nb-O PDF profiles [enlarged part of Fig.~\ref{calc_exp_pdf}(a)].  
(b,c) Contributions to $G(r)$ from pairs oriented in the $ab$ plane and in the $c$ direction, respectively. Numbers in parentheses quantify the amount which a given set of atoms contributes to the total peak. As in Fig~\ref{calc_pdf}(b) thin lines correspond to CG-optimized structure, thick ones to average structure at 800~K. Blue and red lines are for Nb(1) and Nb(2) atoms, respectively.
(d) Temperature evolution of 
Nb displacements from centers of oxygen octahedra
along the $c$ direction 
in time-averaged structures (over 0.2~ps).
The color scale represents an average Nb displacement in eight
octahedra forming a column in the $c$ direction. Red circles mark Nb(1)O$_6$ octahedra.
} \label{NbO}
\end{figure}

We will concentrate now on Nb atoms and their displacements from octahedra centers creating local dipoles. 
Magnified comparison of experimental and calculated distributions [Fig.~\ref{calc_exp_pdf}(a)] of Nb-O bond distances is given in Fig.~\ref{NbO}(a). In Fig.~\ref{NbO}(b,c) we present contributions to this first PDF peak from different Nb-O pairs, taking into account the crystallographically nonequivalent Nb(1,2) and O(1-5) sites.
The difference between the Nb-O profiles involving equatorial O(1,2,3) and apical O(4,5) 
oxygen atoms is obvious at low temperatures [thin lines in Fig.~\ref{NbO}(b) and (c), respectively]:
while for the latter there are two peaks reflecting the tetragonal distortion, the former has a single, rather broad peak. 
There are very substantial differences between Nb(1) and Nb(2) atoms.
The tetragonal distortion coming from displacements of the Nb(1) atoms is bigger and is present even at high temperature, while the Nb(2) only at low temperatures displaces to the extent characteristic for Nb(1) at 800~K [Fig.~\ref{NbO}(c)].  
This smaller shift
of Nb(2) 
along the $c$ axis
is coupled to the
much more off-centered distribution in the \textit{ab} plane 
[see the arrow in Fig.\,\ref{NbO}(b)].
The characteristic shoulder on the side of bigger distances visible in experimental and calculated PDF's can be therefore interpreted as coming both from the tetragonal distortion and from the large in-plane displacements of Nb(2).
%
%

It is interesting to track the development of the octahedral column polarization in time-average structures on lowering temperature.
This is shown in Fig.~\ref{NbO}(d) where the color scale quantifies 
Nb displacements from the center of oxygen octahedron along the $c$ direction
averaged over 8 octahedra in one column. 
In agreement with the partial PDF's, the Nb(1) columns (in red circles) remain strongly
polarized at high temperatures. Additionally one can see that the 
senses of polarization for Nb(1) columns do not change 
within the whole temperature range.
This is not the case for most of the Nb(2) columns, 
for which polarization  
is still dynamic at 420~K and freezes in the temperature range of $420-300$~K. Almost no local polarization change is observed between $300-0$~K.  

\subsection{Vacancies and Ba/Sr disorder}
Finally, we shall complete the analysis of the local distortions by considering the impact of vacancies and Ba/Sr disorder in 
A1 and
A2 channels. Again, it is convenient to follow the  PDF perspective. 
For a calculation of
partial
$G(r)$ distributions, the 'positions' of vacancies have been determined as an average of neighbouring octahedra centers (8 in A1 and 10 in A2 channels). They have been assigned with the Sr scattering length for straightforward comparison with other cations. 

\begin{figure}[!tbp]

\includegraphics[width=90mm,draft=false]{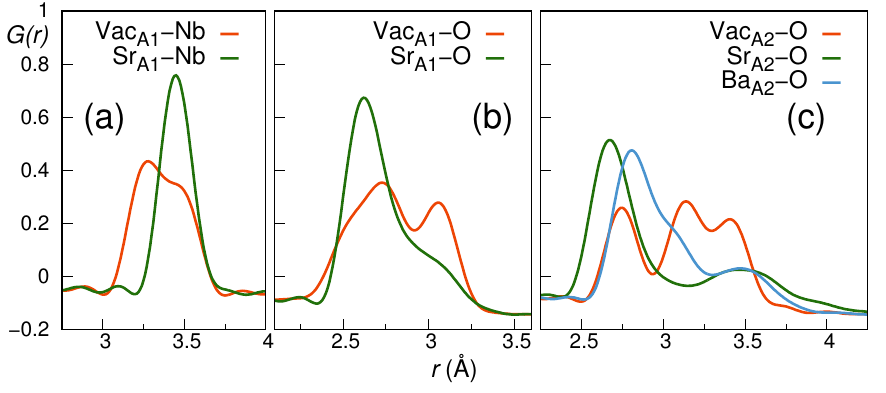}

\includegraphics[width=50mm,draft=false]{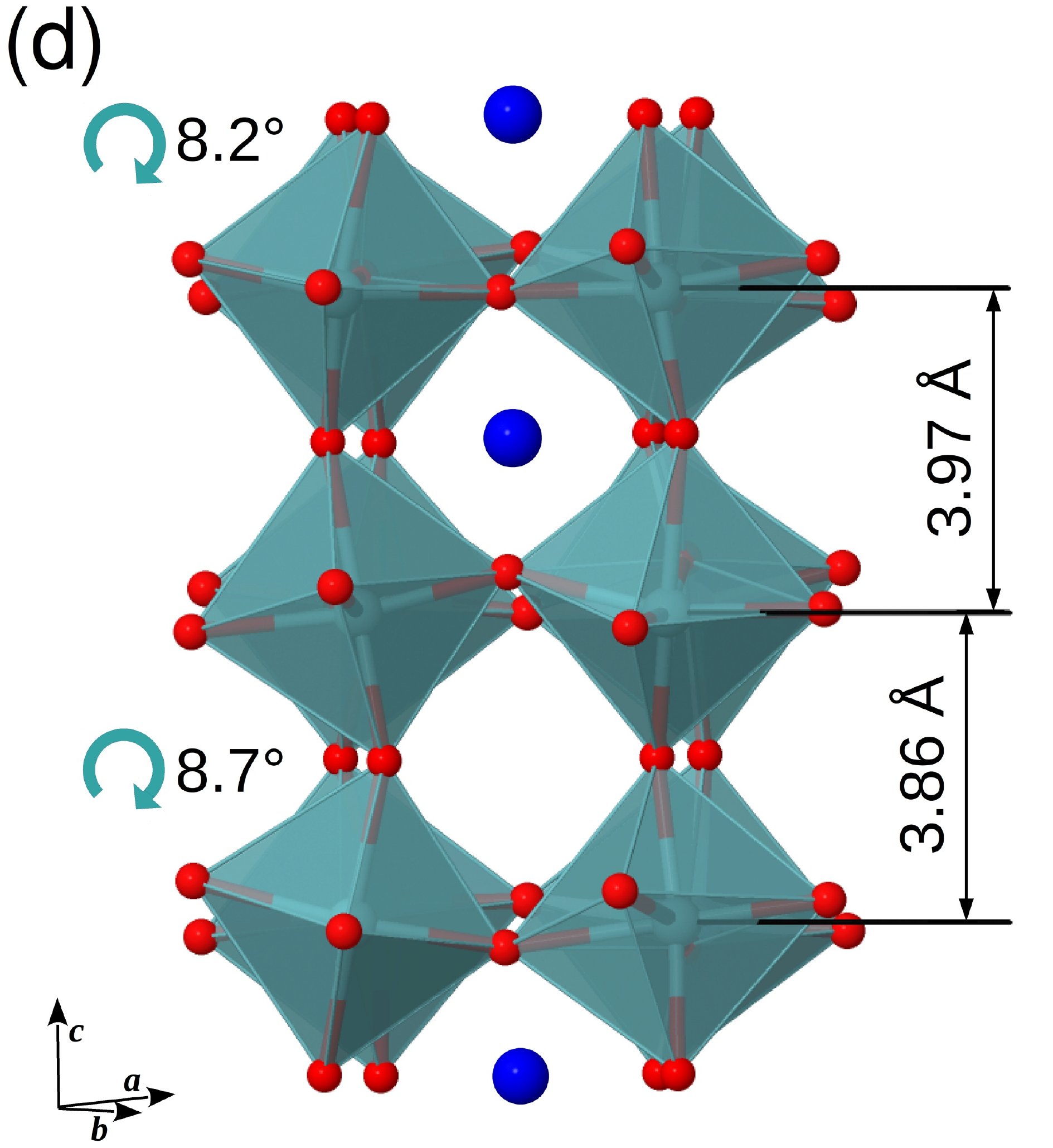}
 \caption{$G(r)$ distributions calculated for pairs including 
Ba/Sr atoms and 
 vacancies in A1 channels -- (a,b) 
 as well as A2 channels -- (c).
 See text for the detailed description of the procedure of calculating distributions including vacancies.
(d) Part of the optimized structure with a vacancy in A1 channel. Average distances between pairs of Nb atoms are given, showing that lack of Sr atom in A1 channel contributes to a substantial local strain. Average tilt for octahedra that do surround a vacancy amounts to 8.7\textdegree, while for those  that do not the angle is 8.2\textdegree. 
} \label{vacancy}
\end{figure}

In Fig.\,\ref{vacancy}(a) we compare 
partial $G(r)$ profiles calculated for the vacancy-Nb(2) and Sr-Nb(2) pairs within A1 channels.
The difference between the two is substantial and cannot be explained just by shifts of Sr atoms adapting to NbO$_6$ polarization (see Appendix, Fig.~\ref{vacancy_supp}).
The double peak of the Vac$_{\rm A1}$-Nb curve with a larger maximum at shorter distances reflects an important local distortion: distances between Nb atoms surrounding a vacancy are $\sim$0.11~\AA{} shorter than the ones observed for a Sr-occupied A1 channel. The impact of a vacancy in the A1 channel is also seen in the $G(r)$ for pairs including nearest oxygen atoms [see Fig.\,\ref{vacancy}(b)]. The information here is two-fold: Sr-O distances are shortened (by octahedra tilts and Sr shifts) and on the other hand a significant peak at larger distances in Vac$_{\rm A1}$-O distribution means that the tilts of octahedra are such that oxygen atoms tend to move away from the center of the channel. Altogether Figs.\,\ref{vacancy}(a,b) produce a picture in which a Nb cube around a vacancy is compressed in $c$ which has to be accompanied by larger tilts of octahedra [see Fig.~\ref{vacancy}(d)]. The tilts are additionally constrained by the O(5) atoms avoiding substantial shifts towards the empty site.

The situation is somewhat simpler in the case of the A2 channels.
From Fig.\,\ref{vacancy}(c) it is clear that both the Ba and Sr atoms deviate from central positions in the channels with the latter displacing more (in accord with the Fig.\,\ref{cg_str}), yielding a maximum of Sr-O distribution at the distances $\sim$0.25~\AA{} shorter. In the Supplementary material [Fig.~\ref{vacancy_supp}(c)] we show $G(r)$ profiles indicating that surrounding octahedra also react more to these larger Sr shifts than in the case of Ba atoms. Vacancies in A2 channels have less tangible impact on the local distortions with no systematic shortening of Nb-Nb distances, as observed in A1 channels. 

\section{Discussion}

From our results, just as from other recent diffuse scattering studies \cite{Pasciak2018,Bosak2015}, it is clear that short-range order in SBN results from an interplay of tilts, polarization and local disorder due to vacancies and Ba/Sr distribution in A2 channels.
Let us first discuss these three components separately.

\subsection{Vacancies, Ba/Sr disorder}

We have seen that the vacancies in the A1 channels act as a local 'negative' pressure bringing Nb(2) atoms closer in the $c$ direction which is accompanied by tilts of larger amplitude. This local strain can be considered as a size-effect~\cite{Butler1992} and as such has been recently identified in x-ray diffuse scattering patterns of SBN35 and SBN81.~\cite{Pasciak2018}
At the same time the vacancies remove screening between apical O(5) atoms and create severe constraints for tilting axes. In that sense one can consider vacancies as  pinning points for octahedra rotations.

In the A2 channels the main effect is related to displacive disorder of Sr and Ba atoms. 
The fact that Sr atoms are more off-center, can be rationalized with substantial underbonding of Sr in A2.~\cite{Woike2003,Graetsch2017} Bond valence sums (BVS) for the average structure~\cite{Podlozhenov2006} that defined initial conditions for our calculations are 1.28 for Sr and 2.01 for Ba. The structure after annealing and optimization has the respective average BVS of 1.83(11) and 2.20(9), meaning that indeed local distortions (displacements and tilts) are such that the Sr BVS is greatly improved. Along with the octahedra tilts, incommensurate modulation of occupancies, displacements and atomic displacement parameters has been proposed for Ba and Sr atoms in pentagonal channels.~\cite{Woike2003,Graetsch2017} 
The size of the supercell limits unfolding of such modulations, nevertheless we observe local modulation of shifts with a period of the doubled unit cell in $c$. This can be clearly seen in Fig.~\ref{zigzag} of the Appendix, where alternation of $ab$ components of Sr/Ba displacements can be observed in some A2 channels. These alternating shifts point to a correlation between tilts and cationic (especially Sr) displacements in A2 channels. Comparison of a pair-distribution profile for Sr-O and its equivalent having Sr moved to the centre of an A2 channel (Fig.~\ref{vacancy_supp} of the Appendix) leads to a similar inference.

\subsection{Octahedra tilts}

Tilts in structures bearing octahedra frameworks have a substantial effect on lattice parameters~\cite{Glazer1972} and provide the means by which the cell volume is optimized (decreased), improving structural stability.~\cite{Reaney1994} Within the SBN family a larger concentration of smaller Sr atoms leads to smaller volumes.~\cite{Paszkowski2013,Podlozhenov2006} It is therefore reasonable to expect, that Sr/Ba ratio
will have its impact on the tilting system. 
Analysis of various experimental results confirms this assumption showing that tilts in Sr-rich concentrations start to appear at higher temperatures (see Fig.~\ref{phase diagram}). 

Our PDF results provide an evidence for differing
tilt development with temperature for the three studied compositions. The 6.2~\AA{} peak splitting (Fig.~\ref{fig2}), which we identify as 
an indicator of the tilt, starts at the highest temperature and reaches the biggest gap at 50~K for SBN61. At 300~K the split has just become visible for SBN35, while it is well developed for SBN61, corresponding to the observation of Schefer et al.~\cite{Schefer2008} who has found a roughly twice larger amplitude of apical oxygen displacements modulation in SBN61 comparing to SBN34.     
%

Experimental PDF's also show that the peak at 2.77~\AA{}, dominated by O-O distances, is noticeably sharp, implying that octahedra are rather rigid. This together with the TTB network provide severe constraints on possible tilt patterns. The existence of triangular octahedra connections limits tilt axes to the $ab$ plane. Whittle \textit{et al.} have conducted a systematic search of possible tilting patterns in a $2^{1/2} \times 2^{1/2} \times 2$ TTB supercell and found just one solution that does not distort octahedra.~\cite{Whittle2015} This solution, however, comprises tilts for octahedra around every second A1 channel, that simultaneously move all apical oxygen atoms towards the centre of the channel. This is certainly unfavorable in the case of vacancies in A1 channels (and also not very probable for filled A1 channels as Sr cations in a non-tilted structure are already overbonded~\cite{Graetsch2017}). The solution that we get from the \textit{ab initio} calculations must, therefore, result from a compromise between an optimum volume and rigid octahedra requirements~\cite{Campbell2018} as well as strong constraints related to vacancies. 

Due to size limits of 
the simulation box we cannot arrive at solutions approximating the incommensurate modulation period observed experimentally. However, it appears from our calculations and from their ability to reproduce experimental PDF's, that the tilting amplitude might be more leveled for all octahedra than available harmonic modulation models of SBN predict. 
Equal tilts also more effectively reduce the volume. 
Altogether the obtained results prompt a picture in which the modulation in SBN is of a statistical character involving highly correlated segments that terminate due to disorder and/or selective tilt correlation 'transmitting' of Nb(1)-based columns.

\subsection{Local dipoles, polarization}

Off centering of Nb atoms in the NbO$_6$ octahedra is considered to be the source of ferroelectric or antiferroelectric behaviour in many compounds including 
perovskite KNbO$_3$ and NaNbO$_3$ or pyrochlore Cd$_2$Nb$_2$O$_7$. Recent DFT calculations have demonstrated that Nb displacements in SBN are stabilized by a partial Nb-O covalency~\cite{Olsen2016} pointing to a standard ferroelectric mechanism. Our calculations give displacements of 0.19~\AA{} for Nb(1) and 0.13~\AA{} for Nb(2) along the $c$ axis at 0~K [\textit{cf.} Fig.\,\ref{NbO}(c,d)]. This can be compared to the room-temperature experimental values for SBN48~\cite{Podlozhenov2006}: 0.15~\AA{} and 0.14~\AA{}, respectively. Other works show similar displacement values.~\cite{Chernaya2000,Graetsch2017}
These off centerings are also comparable to 0.16\,\AA{} observed for KNbO$_3$~\cite{Abrahams1968}. Accordingly,
the macroscopic spontaneous polarization $P_{\rm S}$ is large in all SBN compositions
($0.2-0.35$\,C/m$^2$)~\cite{Glass1969} and is comparable to $P_{\rm S}=0.3$\,C/m$^2$ of KNbO$_3$ \cite{Abrahams1968} and other classical ferroelectrics.~\cite{LinesGlass}  

Our experimental PDF's show that the Nb-O peak changes 
little upon cooling, meaning that similar displacements characterize both high and low temperatures. This is confirmed by ~\textit{ab initio} MD calculations showing that polarized NbO$_6$ chains exist even at the highest temperatures. Many scattering studies of SBN report the observation of planar diffuse intensity perpendicular to \textbf{c*} direction in the reciprocal space which comes from NbO$_6$ chain correlations.~\cite{Gvasaliya2014,Ondrejkovic2014,Ondrejkovic2016,Pasciak2018,Bosak2015,Borisov2013} This polar-chain related diffuse scattering has been shown to display a critical behaviour akin to that of the 
permittivity~\cite{Gvasaliya2014,Ondrejkovic2014}, including characteristic GHz frequencies of dielectric relaxation modes.~\cite{Ondrejkovic2014} From these studies emerges a picture of the phase transition in which polar chains cluster and critically slow down across the $T_{\rm C}$. A detailed x-ray diffuse scattering study revealed still another component of the scattering intensity which indicates that anti-polar arrangements of chains build-up towards the $T_{\rm C}$ as well.~\cite{Bosak2016}

Due to limited simulation times and the polarization correlation length in $c$ exceeding the size of the simulation cell we have limited ability to reproduce the chain clustering and domain formation in our calculations. However, we do observe a change in the dynamics of Nb(2)O$_6$ chains at the temperatures between $420-300$~K  close to the experimental $T_{\rm C}=381$~K~\cite{Paszkowski2013}. A novel and somewhat surprising information is that Nb(1)O$_6$ chains have a very different behaviour. Since they are strongly polarized at high temperatures, 
one can expect them to have a high barrier for the polarization flipping and hence long relaxation times. 
Therefore, when clusters of transversely correlated chains grow towards the phase transition, some of the Nb(1)O$_6$ chains can remain oppositely polarized with respect to their neighbourhood creating anti-polar defects corresponding to the findings of Ref.~\onlinecite{Bosak2016}.

The difference between Nb(1) and Nb(2) has been already seen in former structural studies. However, the reported results are not consistent. For example in the case of SBN61, Nb(1) off centerings in the $c$ direction are larger than Nb(2) ones in Ref.~\onlinecite{Woike2003}, while it is the opposite in Ref.~\onlinecite{Graetsch2017}. On the other hand most of the works predict moderate ($\lesssim$0.05~\AA{}) displacements from octahedra centers within the $ab$ plane for Nb(2) atoms (e.g. Ref.~\onlinecite{Woike2003,Jamieson1968,Graetsch2017}). Not being restricted to any crystallographic model, we get bigger $ab$-plane displacements  $\sim$0.13~\AA{} at 0~K and believe they are one of the key elements of SBN's polar inhomogeneity.

\subsection{Relaxor behaviour}
All these local structure findings give some hints on the origin of the relaxor behaviour for the Sr-rich SBN. 
For small Sr concentrations the unit cell volume is relatively large, tilt is small and polarization develops 'freely' at higher temperatures (with the difference between two types of Nb playing a smaller role). With more Sr atoms tilts start developing at higher temperatures. They reach bigger tilt angles due to both smaller optimal volume as well as bond valence requirements of Sr atoms in A2 channels. As a consequence, local NbO$_6$ dipoles -- mostly those involving Nb(2) atoms -- have larger $ab$-plane components and energy minimum characterizing displacements in the $c$ direction is shallower (cf. displacive mode amplitude and energy from first principles for SBN100~\cite{Olsen2016}). These $ab$-plane components can correlate over short-range distances; indeed, anisotropic in-plane correlations of dipoles have been recently inferred from diffuse scattering in SBN81.~\cite{Pasciak2018}
Furthermore, these in-plane displacements can be also coupled to cation shifts in A2 channels, bigger for Sr atoms, thus having more impact for Sr-rich compositions. In this way the intrinsic disorder of SBN can engage in disrupting long-range polar order in relaxor compositions.

The difference between Nb(1) and Nb(2) can be instrumental in enhancing relaxor properties in Sr-rich SBN. It is reasonable to believe that at temperatures at which the Nb(1)O$_6$ chains freeze, they are very weakly transverse-correlated due to screening. On approaching the $T_{\rm C}$ Nb(2)O$_6$ chains start clustering, but the process is somewhat feeble due to the in-plane turmoil resulting in a smaller polarization of perovskite-like parts of the structure. Therefore, dipolar interactions are too weak to effectively order already strongly polarized Nb(1)O$_6$ chains. This is in accord with huge values of Nb(1)'s atomic displacement parameters U$_{33}$ observed for SBN82 at room temperature
.~\cite{Graetsch2017} The persistence of Nb(1) polarization and Nb(2) partial disorder leads to nanoscale inhomogeneity and relaxor behaviour. The two-subsystem polarization has the potential in explaining some other peculiarities of the relaxor SBN, e.g. its thermal history dependent behavior~\cite{Graetsch2017b,Ondrejkovic2016}, or the two susceptibility peaks upon heating in SBN81.~\cite{Buixaderas2017}





\section{Conclusions}

In conclusion, temperature-dependent neutron pair distribution function experiments were conducted for the first time and were complemented with the first principle calculations. Together these have revealed key ingredients of the local structure of Sr$_x$Ba$_{1-x}$Nb$_2$O$_6$ with $x=0.35, 0.5$ and 0.61, at the same time indicating that the existing structural models are not successful in representing sub-nanometer interatomic distances.
DFT based molecular dynamics calculations were instrumental in separating dynamical contributions to pair distribution profiles and thus unveiling real changes in the structure upon cooling.

The most pronounced temperature dependence of the short-range structure is observed at distances close to 6~\AA{} and it is found to be related to tilting of oxygen octahedra. Clear dependence on Sr content is observed with the local tilting pattern starting to develop at higher temperatures for larger $x$. The tilt 
pattern unfolds as a compromise between the requirements for optimal volume, minimal distortion of octahedra and the constraints from vacancies in square channels. 

These vacancies in square channels are additionally found to be a source of substantial local strain, shortening distances between coordinating Nb atoms by $\sim$0.1~\AA{} in the $c$ direction. 

Nb atoms are 
displaced from centers of oxygen octahedra at high temperatures and the distribution of displacement distances is preserved in the low-temperature structure. Different behavior on the two distinct crystallographic positions of Nb is observed from \textit{ab initio} calculations with Nb(1) atoms shifting mostly along the tetragonal axis and Nb(2) atoms having large \textit{ab}-plane components. 
Nb(1)O$_6$ chains are characterized by local polarization that 
freezes
at high temperatures, well above the 
$T_{\rm C}.$
On the contrary, polarization of Nb(2)O$_6$ chains 
ceases to fluctuate at temperatures close to $T_{\rm C}$. 
We argue that the $ab$-plane
displacement disorder of Nb(2) atoms, interconnected with  octahedra tilts and Sr displacements, 
is responsible for relaxor behaviour of Sr-rich SBN.

\begin{acknowledgments}
 This work was supported by the Czech Science Foundation (project no. 16-09142S). The computational part of this research was undertaken on the NCI National Facility in Canberra, Australia, which is supported by the Australian Commonwealth Government. 

\end{acknowledgments}

\section*{Appendix}

\begin{figure}[tbh]
\includegraphics[width=70mm,draft=false]{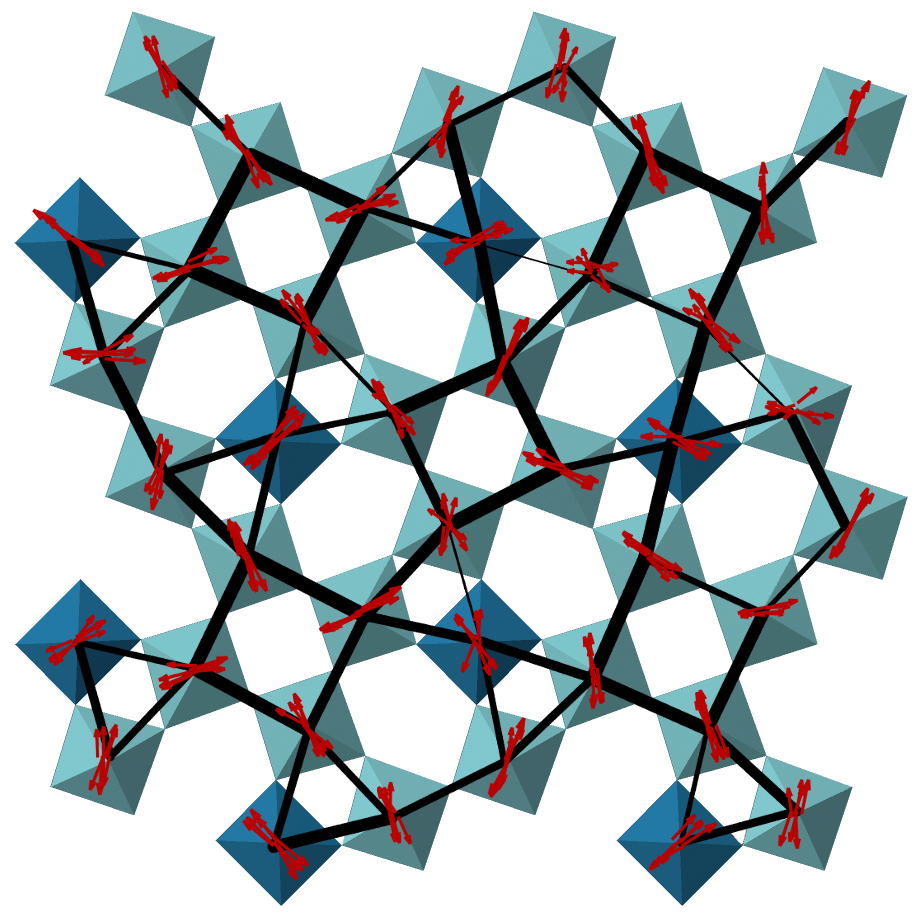}
 \caption{Similar as Fig.~\ref{tilt}(b) but the tilt vectors are not integrated to one plane, which serves showing that the dispersion of vector directions within one column is usually very small. Vector senses alternate from one layer to the next one.
} \label{tilt_supp}
\end{figure}

\begin{figure}[tbh]
\includegraphics[width=70mm,draft=false]{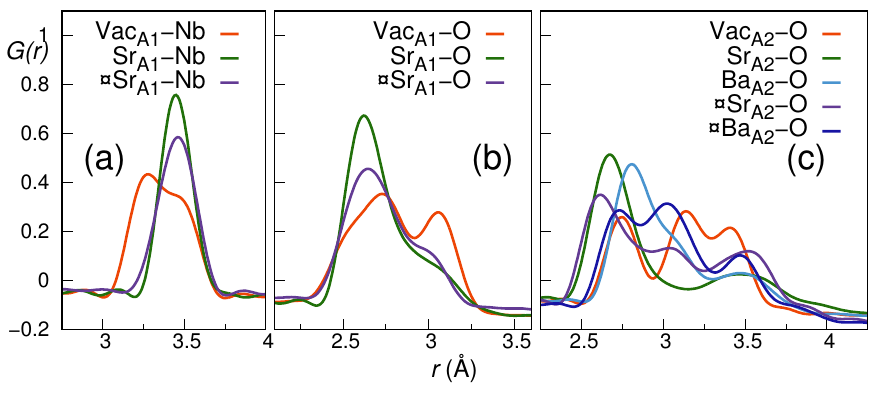}
 \caption{Similar as Fig.~\ref{vacancy}. Additional profiles marked with \textcurrency{} are produced with respective cations placed in positions calculated as a geometrical average of centers of surrounding octahedra (just as in the case of vacancies). In this way these distributions have an information on a given cation's neighbourhood $G(r)$ irrespectively of the cations displacement. This is useful as it allows one for several additional observations. The difference between Sr$_{A1}$-Nb and \textcurrency{}Sr$_{A1}$-Nb distributions in (a) comes from the fact that the Nb atoms shift from the centers of octahedra (broad peak of \textcurrency{}Sr$_{A1}$-Nb), a movement to which neighbouring Sr atoms adapt (sharper Sr$_{A1}$-Nb peak). With this differentiation it is possible to exclude polarization as a source of the large split of the Vac$_{A1}$-Nb peak leaving local strain as the only suspect. Similarly, an enhanced peak at higher distances for Vac$_{A1}$-O in (b) can be attributed solely to the 'repulsive' effect of the lack of screening (see main text). There is interesting information in a somewhat busy graph (c) as well: from the fact that Sr$_{A2}$-O and \textcurrency{}Sr$_{A2}$-O distributions are much more alike than their Ba counterparts one can deduce that octahedra network reacts more to Sr displacements than to Ba ones.      
} \label{vacancy_supp}
\end{figure} 

\begin{figure}[tb]
\includegraphics[width=70mm,draft=false]{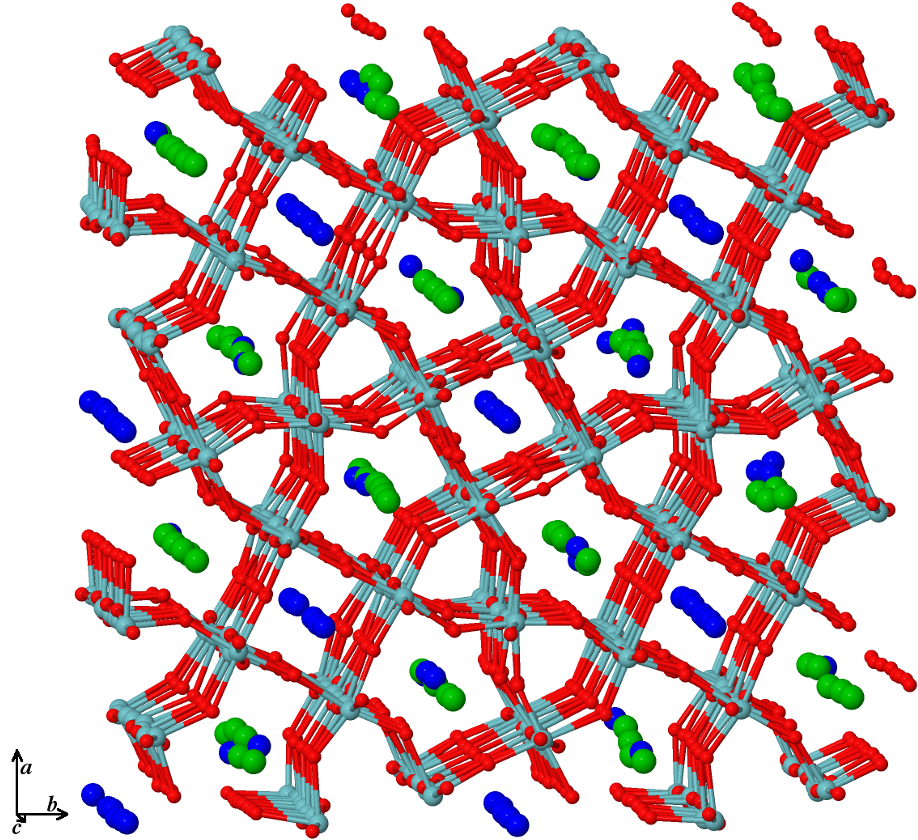}
\caption{Yet another view on the optimized structure. This one is to show that zig-zag displacements of Sr and Ba atoms are present in A2 channels.      
} \label{zigzag}
\end{figure}

\end{document}